\documentclass[12pt]{article}
\pdfoutput=1
\usepackage{amsmath,amssymb,amsfonts,amsthm,epsfig,url,color,todonotes}

\usepackage{tikz}
\usepackage{multirow}\usepackage{multicol} 

\newcommand{\comment}[1]{}

\newcommand{\IF}{\mathbb{F}}
\newcommand{\IR}{\mathbb{R}}
\newcommand{\cN}{\mathcal{N}}

\begin{document}

   \begin{center}
{\Large \bf Machine-learning a virus assembly fitness landscape}
\end{center}
\

\vspace{.4cm}
\centerline{
 {
  {\large Pierre-Philippe Dechant}$^1$ and
  {\large Yang-Hui He}$^2$
}}
\vspace*{3.0ex}

\begin{center}
  {\it
    {\small
${}^{1}$
      Pro Vice Chancellor's Office, \\York St John University, York, YO31 7EX, and \\
    York Cross-disciplinary Centre for Systems Analysis\\
	University of York, Heslington YO10 5GE, UK\\
          {\rm \url{ppd22@cantab.net}}\\
    }
\vspace*{1.5ex}
    {\small
      ${}^{2}$ 
        Dept.~of Mathematics, City, University of London, EC1V 0HB, UK; \\
        Merton College, University of Oxford, OX14JD, UK and\\
         School of Physics, NanKai University, Tianjin, 300071, P.R.~China \\
        \qquad
            {\rm \url{hey@maths.ox.ac.uk}}\\
    }
  }
\end{center}

\vspace{1in}

\begin{abstract}
\comment{
Evolutionary fitness landscapes are notoriously difficult to construct. We build on a recent model of virus assembly that captures key aspects of the assembly mechanism, which takes assembly efficiency as a proxy for fitness. This model uses cutting-edge insights into the cooperative role of multiple  packaging signals dispersed throughout the viral genomic RNA, which act as a nucleation site for assembly  and help recruit capsid protein onto the growing capsid. This model consists of a dodecahedral capsid with $12$ corresponding packaging signals in three affinity bands. This whole $3^{12}$ genome space has been explored via stochastic assembly models, giving a fitness landscape in terms of the assembly efficiency. These stochastic simulations are computationally very costly; however, the data format of a 12-dimensional vector being mapped to a number  lends itself to an artificial intelligence approach.  Here we have trained a neural network on a subset of the data to predict the remaining data to high accuracy. This machine-learning approach might lead the way to more complicated models through bootstrapping of an artificial intelligence by only partially exploring the genome space for training purposes, and then predicting fitness procedurally as necessary. }

Realistic evolutionary fitness landscapes are notoriously difficult to construct. A recent cutting-edge model of virus assembly consists of a dodecahedral capsid with $12$ corresponding packaging signals in three affinity bands. This whole genome/phenotype space consisting of $3^{12}$ genomes has been explored via computationally expensive stochastic assembly models, giving a fitness landscape in terms of the assembly efficiency.
Using latest machine-learning techniques by establishing a neural network, we show that the intensive computation can be short-circuited in a matter of minutes to astounding accuracy.
\end{abstract}

\newpage


\date{\today}

\section{Introduction}
Two facts about simple viruses have been known for a long time. Firstly, that genetic economy leads to the use of symmetry, such that virus capsids are mostly icosahedral or helical. Secondly, packaging signals, that is secondary structure features in the viral RNA, are often required for encapsidation in viruses with single-stranded genomes. Examples are the origin of assembly sequence in Tobacco Mosaic virus, the psi element in HIV and the TR sequence in MS2. This is an evolutionary advantage, as it ensures vRNA-specific encapsidation and can increase assembly efficiency through a cooperative role of the RNA, which acts as a nucleation site. 

\begin{figure}
	\begin{center}
		\includegraphics[height=7cm]{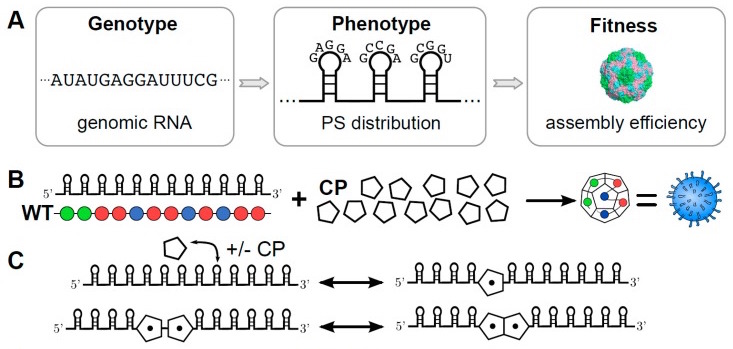}
\caption[fourD]{\textbf{A} The nucleotide sequence of a virus determines the gene products; however, in addition to this information content the RNA also explores a configuration space of secondary structures. Viruses appear to have evolved to use such motifs to help recruit coat protein with a conserved common recognition motif during assembly. The stability and binding affinity of these packaging signals gives a distinctive profile for viral assembly, which is the phenotype relevant to assembly. Assembly efficiency is the fitness of this phenotype, or at least the contribution to the overall fitness that is determined by aspects of assembly.   \textbf{B} The genomes in the model consist of twelve packaging signals (PS) that can take weak, medium and strong binding affinities. They successively recruit twelve pentagonal coat proteins, which together form the dodecahedral virion in this model. \textbf{C} The stochastic simulation algorithm models several possible reactions. Firstly, packaging signals can bind coat proteins (and fall apart again), and secondly, two coat proteins that have been recruited by packaging signals can bind to each other. The fitness landscape was computed for 2000 virions for each possible genome, making the computations very intensive. }
\label{figPSMA}
\end{center}
\end{figure}

More recently, it has been shown that taken together, these two facts suggest that there could be more than one packaging signal, with multiple signals in fact dispersed throughout the genome. This is because the capsid is symmetric, and the packaging signal mechanism functions via interaction between viral RNA and the coat protein (CP). In several cases, this RNA-CP interaction leads to a conformational change in the CP, which only then makes it assembly competent (e.g. TMV and MS2). The picture that emerges is then that there are multiple packaging signals (PS) that recruit CP onto a growing capsid. This reduces the phase space that CP has to search in order to assemble a capsid, resulting in vastly increased assembly efficiency. The details of such a mechanism were found in MS2 and STNV in 
 \cite{rolfsson2016direct,Rolfsson:2010,dykeman2014solving,Morton:2010,stockley2016bacteriophage,stockley2013packaging,dykeman2013packaging,stockley2013new,dent2013asymmetric,koning2016asymmetric,geraets2015asymmetric,patel2015revealing,twarock2018RNA,twarock2018hamiltonian}. 
Once the details of this mechanism were understood using biochemistry, structural biology, bioinformatics, biophysics and graph theory in these model systems, related mechanisms could be found in clinically relevant viruses
\cite{shakeel2017genomic,stewart2016identification,patel2017hbv}. These packaging signals are secondary structure features of the viral genomes where a stemloop in the single-stranded RNA presents a common recognition motif that can bind to CP (see Fig. \ref{figPSMA}\textbf{A}). The viral genome thus has multiple layers of constraints, by having to code for genes as well as the PS instruction manual. This set of packaging signals can also be repurposed and optimised for the assembly of virus-like particles, which do not share the same genetic constraints as the virus, and could be used e.g. for vaccines, drug delivery or as an anti-viral strategy \cite{patel2017rewriting,dechant2019DIPs}. 

An equilibrium model of how to assemble a simplified example of an icosahedral virus, a dodecahedron built from 12 pentagonal faces, was considered in
\cite{zlotnick1994build} using ODEs. 
More recently, the multiple dispersed packaging signal paradigm has sparked renewed interested in such a dodecahedral model \cite{dykeman2014solving,dykeman2013building}. 
The assembly reaction kinetics was modelled via a  set of discrete reactions in a stochastic simulation paradigm based on the Gillespie algorithm
\cite{gillespie1977exact}. 

In this model $12$ PSs can bind CP, as well as dissociate again, reflecting reversible/equilibrium kinetics. Bound CP can then bind other bound CP, gradually building up a capsid (see Fig. \ref{figPSMA}\textbf{B} and \textbf{C}). The PSs here have three different bands of binding affinity: weak, medium and strong. These choices correspond to binding energies of 4/8/12 kcal/mol respectively, based on the TR sequence in MS2 which has approximately 12kcal/mol. The binding energy between CPs is much lower, at approximately 2 kcal/mol. This modulation of affinity affects the assembly kinetics, e.g. by providing a nucleation point that starts assembly, or allowing for error-correction via weaker binding elsewhere. The thermodynamics of PS binding and of the number of CP bonds formed then translates into assembly efficiency. This in turn is taken as a proxy for fitness (all other things being equal) -- or at least the contribution to the fitness that results from assembly considerations \cite{dykeman2017model}.

In \cite{twarock2018modelling} the whole space of these $3^{12}$ genomes (or rather, phenotype profiles) has been explored. The assembly efficiency there is given by the number of capsids that have correctly assembled out of a possible total of $2,000$. This efficiency provides a fitness landscape on the 12-dimensional genome space. This is an interesting model that is tractable, in contrast with many other biological systems, as it has a small number of degrees of freedom and is dominated by the symmetry of the capsid. This tractability also allows for the consideration of viral evolution. For instance, mutation of the PS strengths leads to the emergence of a set of related genomes that form a `quasispecies' \cite{eigen2000viruses}. One can thus investigate the effect of evolutionary pressures, e.g. those exerted by standard drugs or a novel type of drug that targets packaging signals \cite{bingham2017rna}. 

This model thus captures many interesting aspects of viral genetics, geometry and assembly. A more realistic model would have more CP building blocks and PSs, e.g. around $60$ for MS2 (i.e. one full orbit of the icosahedral group). But the computation time for even these simple genomes and the assembly kinetics that provide the fitness landscape are already considerable. Even other simplified models, e.g. reduced orbits  on symmetry axes given by e.g. an icosahedron consisting of 20 triangles with 20 PSs, a rhombic triacontahedron consisting of 30 rhombuses with 30 PSs, or a finer gradation of binding affinity bands are already computationally out of reach.

However, this data set is a perfect example of data that is amenable to a machine-learning approach, since it associates a vector input with a number output. We therefore train a neural network to predict the fitness landscape. The network is trained on a subset of the whole genome space, and validated on the remainder of the data. This proof-of-principle shows that it is very fast for a neural network to learn the inherent patterns within the large degeneracy of the detailed stochastic modelling to predict assembly efficiency fitness for unseen genomes. The danger is that some subtleties of the stochastic modelling are lost, but allowing for  computation times many orders of magnitude faster. This approach could thus in future be used to tackle more realistic models such as the ones mentioned above. Stochastic simulations could be used to partially explore these larger genome spaces, calculating assembly fitness in order to provide a training set for a neural network. The rest of the fitness landscape can then be predicted by the artificial intelligence; it is also possible to only compute this fitness if necessary, e.g. when a new genome arises through mutation in a quasispecies model, such that such computation may only be necessary `precedurally'.

\section{Methodology and Results}
From a purely mathematical point of view, we have the following problem.
Let (weak, medium, strong) be denoted respectively by $(1,2,3)$. The input is a vector $v$ in a 12-dimensional vector space over  $\IF_3$, the field of three elements. 
The output is an integer (which we treat as a real number) between $0$ and $2000$, which we can normalise into $\epsilon \in [0,1] \subset \IR$ by dividing by $2000$.
The algorithm used by 
\cite{dykeman2014solving,twarock2018modelling}
 is thus a map
\begin{equation}
    f \ : \ v \in \IF_3^{12} \longrightarrow \epsilon \in [0,1] \ .
\end{equation}
A typical example is
\begin{equation}\label{eg}
    \{1,1,1,2,2,2,3,1,2,2,1,1\} \longrightarrow \frac{1523}{2000} \simeq 0.7615
\end{equation}

The map $f$ is a computationally intensive one 
(with individual genome run times between 20 minutes and 12 hours, and cumulative run time of 3-4 weeks on the N8 Polaris high performance computing research cluster, Intel 2.6 GHz Sandy Bridge E5-2670 processors, with a total of 5,312 cores,  with a mix of 4 and 16Gb of RAM, https://n8hpc.org.uk/facilities/).
Nevertheless a brute-force simulation has been performed on the $3^{12} = 531,441$ possible input values and the efficiency value extracted.
This gives us a database of some half a million known cases of the form \eqref{eg}.

Such a problem is perfectly adapted to supervised machine-learning: we know many input values and wish to train some artificial intelligence to associate the input with the known output on some small subset, and use it to predict the output for unseen input.
The advantage of this approach is that often approximate results can be attained at reduction in computation time by many orders of magnitude.
The paradigm of using machine-learning in algebraic geometry and more general classes of problems in pure mathematics was proposed in \cite{He:2017aed,He:2017set,He:2018CYbook} to satisfying accuracy, and it is a similar philosophy that we will adopt here.

\begin{figure}[t!!!]
	\begin{center}
		\includegraphics[trim=0mm 30mm 50mm 0mm, clip, width=6in]{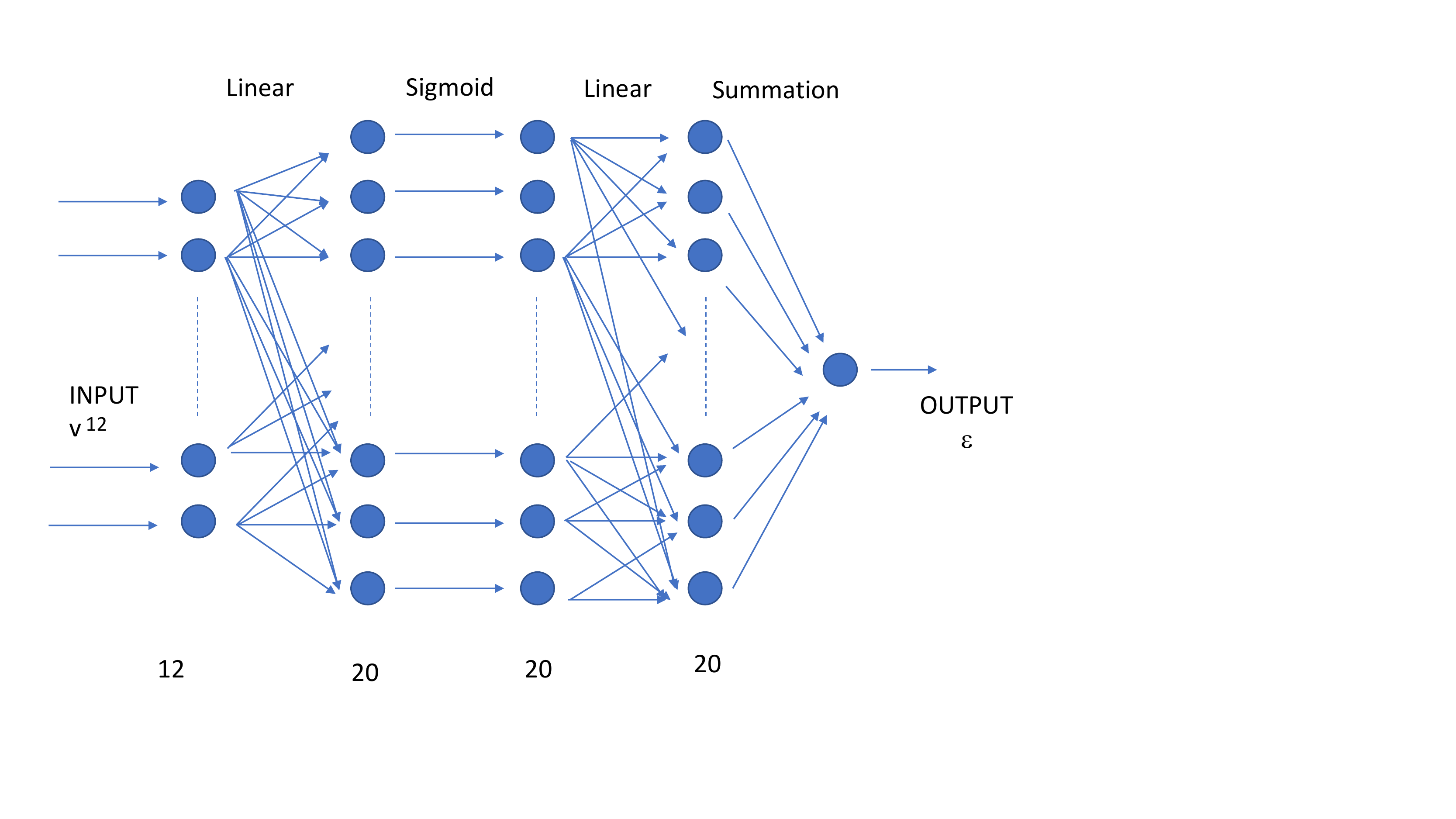}
\caption[fourD]{The structure of the neural network used in the calculation. }
\label{f:NN}
\end{center}
\end{figure}

Let us first try the following specific procedure:
\begin{itemize}
    \item Take the full data $D$ of the form \eqref{eg}, of size $3^{12}$;

    \item Establish the neural network $\cN$, a 3-layer perceptron

    $$\boxed{\text{INPUT} = v}
    \longrightarrow \boxed{L_{20}} 
    \longrightarrow
    \boxed{S_{20}}
    \longrightarrow
    \boxed{L_{20}}
    \longrightarrow
    \boxed{\Sigma_{1}} 
    \longrightarrow
    \boxed{\text{OUTPUT} = \epsilon }$$

    In the above, L means a linear-layer, S, a sigmoid layer and $\Sigma$, a summation layer.
    In particular,
    the first linear layer L$_{20}$ is a fully connected layer taking the 12-vector $v$ to 20 neurons by simply the linear function $y =  w x + b$.
    This is then  fed into an element-wise sigmoid layer $\sigma(x) = (1+e^{-x})^{-1}$ of 20 neurons, followed again by a linear layer, which is then summed to the real number $\epsilon$ as the output.
    The schematic of $\cN$ is shown in Figure \ref{f:NN}.
    We have taken this neural network only to illustrate the power of our methodology and have not optimised the hyper-parameters such as $20$, nor the network architecture or the choice of the type of neurons.
    \item Now split $D$ into a training set $T$ of 30,000 random samples; the validation set will be the complement $V = D \backslash T$. Note that the training data is only about 5.6\% of the total data.
    
    \item We train $\cN$ with $T$ and validate on $V$.
    
    \item As a further check, we create a ``fake'' validation set $\widetilde{V}$ which has the same inputs as that of $V$ but with output randomly assigned from the set of correct outputs.
\end{itemize}


\begin{figure}[th!!!]
\textbf{A} 
\includegraphics[trim=0mm 0mm 0mm 0mm, width=2.5in]{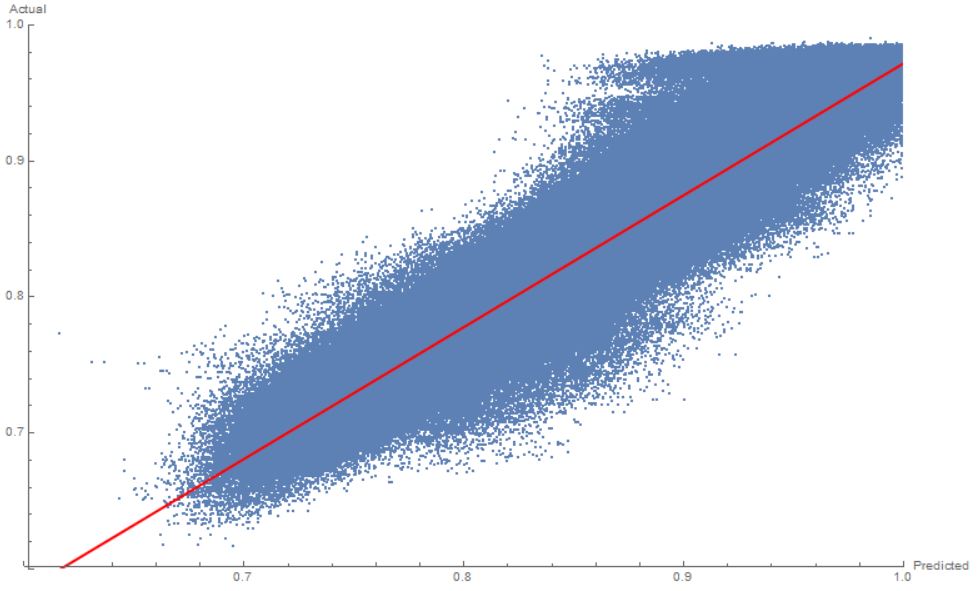}
\textbf{B} 
\includegraphics[trim=0mm 0mm 0mm 0mm, width=2.5in]{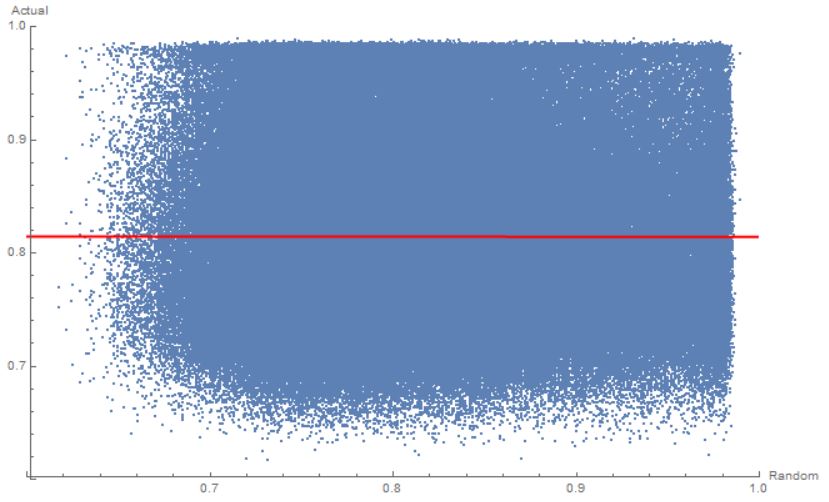}
\caption{
{
\textbf{A}  A scatter plot of the predicted and actual value of $\epsilon$ for the validation data from a $30,000$ training sample; \textbf{B}  scatter plot of a random-prediction versus the actual $\epsilon$ value.
}
\label{f:30K}}
\end{figure}

On an ordinary laptop (Intel Core m5-6Y57 CPU,  1.10GHz, 2 Cores, 4 Logical Processors, with 8Gb of RAM), the training took about 45 seconds, and the prediction about 10 seconds.
The algorithm is implemented on Mathematica \cite{mathematica} and is expected to even faster on the Python package Tensorflow \cite{tensorflow}.
In other words, the entire computation took under 1 minute with an ordinary notebook as opposed to the many hours it took on a supercomputer.


We present the result in Figure \ref{f:30K}.
In part \textbf{A}, we present a plot of the predicted $\epsilon$ on the horizontal versus the actual $\epsilon$ on the vertical. There are $501,441$ points.
One can see that they cluster near the desired $y=x$ line, which would mean perfect prediction (note the axis ranges).
To give some precise measures, the best fit line is $y = -0.0262122+1.02519 x$ with F-statistic $2.45082 \times 10^6$ and p-value less than $10^{-10^6}$.
The R-squared value is $0.830151$.
To double check, we plot the same result for the fake validation set $\widetilde{V}$ in part \textbf{B}.
It is obvious that the distribution is much less structured and essentially randomly occupies a square.
The fit here is $y=0.815233 -0.000947474 x$ i.e. practically a constant, with a poor F-statistic of
$0.450401$ and a poor p-value of $0.502145$.
The R-squared value is $8.98217 \times 10^{-7}$.
This is very re-assuring for less than 6\% of seen data and total computation time of less than 1 minute on an ordinary laptop.

\begin{figure}[th!!!]
\includegraphics[trim=0mm 0mm 0mm 0mm, width=6in]{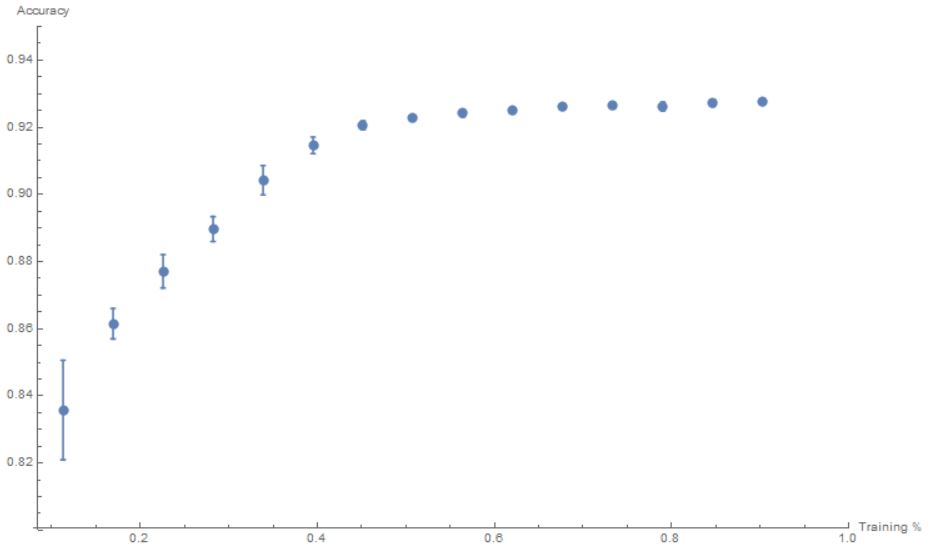}
\caption{{Learning curve for the R-square value versus fraction (seen) training data from 0 to 1.
}
\label{f:learningC}}
\end{figure}

To give an idea of the prediction, for the $(1,1,\ldots,1)$ vector, the net predicts
$\epsilon = 0.87069$, or $1741$. The original value is $200$, but that is a singular outlier in the whole data set, which we would not expect the neural network to be able to reproduce. 
For the $(2,2,\ldots,2)$ vector, the net predicts
$\epsilon = 0.834721$, or $1669$; the correct value is $1745$.
For the $(3,3,\ldots,3)$ vector, the net predicts
$\epsilon = 0.673568$, or $1347$; the correct value is $1309$.

We will use R-squared, a real number between 0 and 1, as a measure of accuracy of the machine-learning; the closer it is to 1, the better the fit (for a good reference on machine-learning and goodness of fit measure, cf.~e.g.\cite{NN}).
We present the {\bf learning curve} for R-squared to demonstrate the response of the neural network in Figure \ref{f:learningC}.
We split the data into a fraction $x$ of random samples for training and validate on the complement $1-x$, done for training set from $30,000$ to $500,000$, in steps of $30,000$.
The R-squared value is computed for each case  as a measure of precision.
For each $x$, we repeat the random sampling 10 times, for which we get the error bars.

As a comparison, one might imagine that since there is an underlying pattern being machine-learnt, a simple regression might suffice.
That is, could one fit a hyperplane $f(x_i) = a_0 + \sum\limits_{i=1}^{12} a_i v_i$ to the data? We perform this over the entire dataset, and find that the best multi-linear regression obtains only $R^2 = 0.575428$.
Introducing non-linearity and more parameters, such as fitting $f(x_i) = a_0 + \sum\limits_{i=1}^{12} a_i v_i + \sum\limits_{i=1}^{12} b_i v_i^2$ does not do much better, at $R^2 = 0.665484$. The inherent complexity (non-linearity) of the problem is therefore best captured by our neural network approach. 




\subsection*{Ethics statement}
This work did not use personal data, nor does the research have ethical implications. The research was performed with the highest standards of academic integrity. 
\subsection*{Data accessibility}
This work does not have any experimental data. 
\subsection*{Competing interests}
The authors have no conflicts of interest to declare.
\subsection*{Author's contributions}
Both authors contributed equally to the paper. 
\subsection*{Acknowledgements}\label{ack}
We would like to thank Richard Bingham and Eric Dykeman for making available their code, fitness landscape and figures. We would like to thank Reidun Twarock, Richard Bingham and Eric Dykeman for interesting discussions. 
YHH is indebted to the STFC for grant ST/J00037X/1.

\subsection*{Funding statement}
This work was not funded by any specific grant.

\bibliography{virobib}

\end{document}